\documentclass[aps,preprint,showpacs]{revtex4}
\usepackage[T1]{fontenc}
\usepackage[latin1]{inputenc}
\usepackage{amsmath}
\usepackage{amssymb}
\usepackage{graphicx}
\usepackage{esint}

\makeatletter
\@ifundefined{textcolor}{}
{%
 \definecolor{BLACK}{gray}{0}
 \definecolor{WHITE}{gray}{1}
 \definecolor{RED}{rgb}{1,0,0}
 \definecolor{GREEN}{rgb}{0,1,0}
 \definecolor{BLUE}{rgb}{0,0,1}
 \definecolor{CYAN}{cmyk}{1,0,0,0}
 \definecolor{MAGENTA}{cmyk}{0,1,0,0}
 \definecolor{YELLOW}{cmyk}{0,0,1,0}
}

\makeatother

\begin{document}

\title{Collective mode evidence of high-spin bosonization in a trapped one-dimensional
atomic Fermi gas with tunable spin}

\author{Xia-Ji Liu$^{1}$ }

\email{xiajiliu@swin.edu.au}

\author{Hui Hu$^{1}$}

\email{hhu@swin.edu.au}

\affiliation{$^{1}$Centre for Quantum and Optical Science, Swinburne University
of Technology, Melbourne 3122, Australia}

\date{\today}
\begin{abstract}
We calculate the frequency of collective modes of a one-dimensional
repulsively interacting Fermi gas with high-spin symmetry confined
in harmonic traps at zero temperature. This is a system realizable
with fermionic alkaline-earth-metal atoms such as $^{173}$Yb, which
displays an exact SU($\kappa$) spin symmetry with $\kappa\geqslant2$
and behaves like a spinless interacting Bose gas in the limit of infinite
spin components $\kappa\rightarrow\infty$, namely high-spin bosonization.
We solve the homogeneous equation of state of the high-spin Fermi
system by using Bethe ansatz technique and obtain the density distribution
in harmonic traps based on local density approximation. The frequency
of collective modes is calculated by exactly solving the zero-temperature
hydrodynamic equation. In the limit of large number of spin-components,
we show that the mode frequency of the system approaches to that of
a one-dimensional spinless interacting Bose gas, as a result of high-spin
bosonization. Our prediction of collective modes is in excellent agreement
with a very recent measurement for a Fermi gas of $^{173}$Yb atoms
with tunable spin confined in a two-dimensional tight optical lattice. 
\end{abstract}


\maketitle

\section{Introduction}

Ultracold atomic gases appear to be a versatile tool for discovering
new phenomena and exploring new horizons in diverse branches of physics.
To large extent, this is due to their unprecedented controllability
and purity. A vast range of interactions, geometries and dimensions
is possible: using the tool of Feshbach resonances \cite{FR} and
applying a magnetic field at the right strength, one can control very
accurately the interactions between atoms, from arbitrarily weak to
arbitrarily strong. By using the technique of optical lattices that
trap atoms in crystal-like structures \cite{OpticalLattice}, one
can create artificial one- or two-dimensional environments to explore
how physics changes with dimensionality. Most recently, it is also
able to control the number of spin-components: degenerate Fermi gas
with high-spin symmetry has been observed in alkaline-earth-metal
atoms Yb \cite{Fukuhara2007,YbLENS}. 

The ytterbium (Yb) atom has a unique advantage in studying high-spin
physics. It has a closed-shell electronic structure in the ground
state, and hence its total spin is determined entirely by the nuclear
spin, $I$. For the fermionic species $^{173}$Yb, the nuclear spin
is $I=5/2$ and the atom can be in six different internal states $\kappa=2I+1=6$.
This gives rise to a unique feature of $^{173}$Yb atom, that is,
one simple internal-state-independent $s$-wave scattering length
due to the absence of electronic spin in the atomic ground state \cite{Kitagawa2008}.
Thus, the system exhibits SU(6) symmetry \cite{Hermele2009,Cazalilla2009,Gorshkov2010}.
In such a high-spin system, possible novel ground states and topological
excitations have been addressed theoretically \cite{Hermele2009,Cazalilla2009,Gorshkov2010,Wu2003}.
One dimensional (1D) repulsively interacting fermions with sufficiently
high spin also behave like spinless interacting Bose atoms \cite{Yang2011},
a phenomenon that we may refer to as \emph{high-spin bosonization}.
Physically, this phenomenon may also occur in two or three dimensions.

\begin{figure}
\begin{centering}
\includegraphics[clip,width=0.8\textwidth]{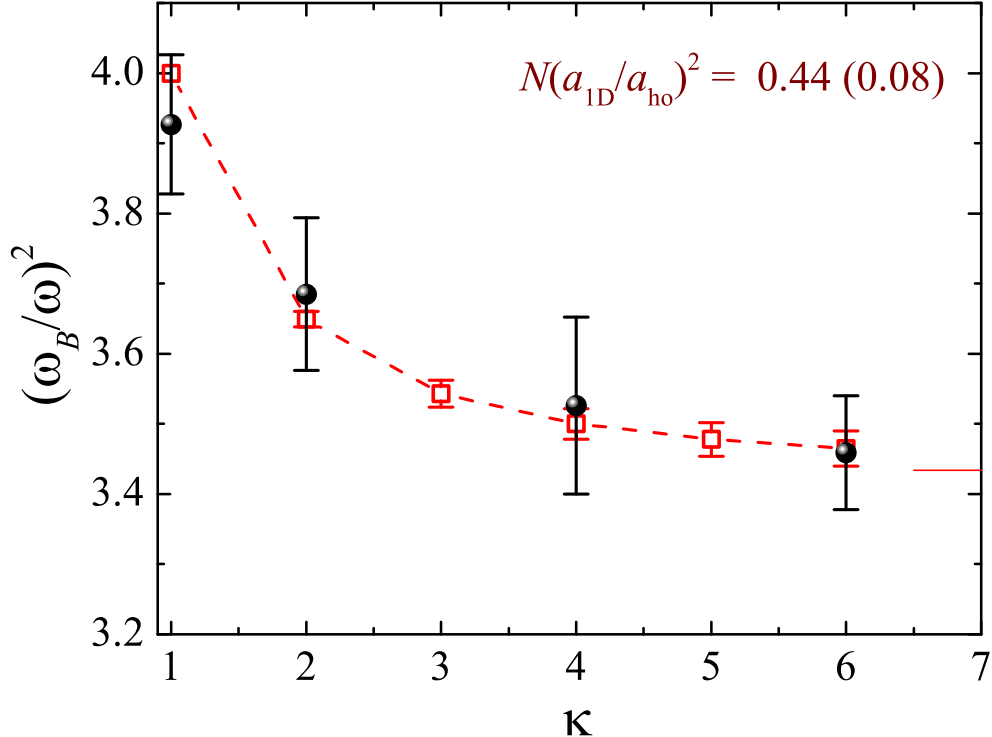} 
\par\end{centering}

\caption{(Color online) Breathing mode frequency as a function of the number
of components at the dimensionless interaction parameter $Na_{1D}^{2}/a_{ho}^{2}=0.44\pm0.08$.
The solid circles with error bars are the experimental data reported
by the LENS team \cite{YbLENS}. The empty squares are the theoretical
results, with error bars counting for the experimental uncertainty
for interaction parameter. The thin horizontal line at the right part
of the figure shows the theoretical prediction at the infinitely large
number of components. \textit{Source}: Adapted from Ref. \cite{YbLENS}.
$ $Copyright 2014, by Nature Publishing Group.}

\label{fig1} 
\end{figure}

Experimentally, Fermi degeneracy of $^{171}$Yb and $^{173}$Yb atoms
has been demonstrated \cite{Fukuhara2007,Taie2010,YbLENS}. In particular,
in a very recent experiment performed at European Laboratory for Non-Linear
Spectroscopy (LENS), a Fermi gas of $^{173}$Yb atoms has been created
in one-dimensional harmonic traps by using optical lattices and its
momentum distribution and breathing mode oscillation have been measured
\cite{YbLENS}. 

In this work, motivated by the recent measurement at LENS \cite{YbLENS},
we investigate a 1D high-spin Fermi gas with strongly \emph{repulsive}
interactions satisfying SU($\kappa$$ $) symmetry, by using exact
Bethe ansatz technique beyond the mean-field framework \cite{Sutherland1968}.
First, we solve the exact ground state of a homogeneous Fermi gas
at zero temperature based on Bethe ansatz. To make contact with the
experiment, we then consider an inhomogeneous Fermi cloud under harmonic
confinement, within the framework of local density approximation (LDA).
The equation of state of the system and the density distribution are
calculated. By solving the zero-temperature hydrodynamic equation,
we predict the frequency of low-lying collective modes. The main outcome
of our research is summarized in Fig. \ref{fig1}, which shows the
predicted breathing mode frequency in comparison with the recent data
reported by LENS group, for a trapped $^{173}$Yb gas at a dimensionless
interaction parameter $Na_{1D}^{2}/a_{ho}^{2}=0.44\pm0.08$ \cite{YbLENS}.
Here, $N$ is the total number of atoms in the Fermi cloud, $a_{1D}$
and $a_{ho}$ are the 1D \textit{s}-wave scattering length and the
length of harmonic oscillator, respectively. We find an excellent
agreement between our theoretical prediction and the experimental
data, with a relative discrepancy at a few percents. Fig. 1 has been
published in Ref. \cite{YbLENS}. The purpose of this paper is to
present the details of our calculations, focusing particularly on
the high-spin bosonization.

We note that in earlier works 1D multi-component Fermi gases with
large-spin and attractive interactions have been discussed in detail
\cite{amulti1,amulti2}. However, for repulsive interactions, only
homogeneous Fermi gas in weak and strong coupling limits has been
considered \cite{rmulti}. Comparing with the case with attractive
interactions there are no multi-component bound clusters in repulsively
interacting Fermi gases. 

The paper is organized as follows. In the following section, we describe
briefly the model Hamiltonian. In Sec. III, we present the exact Bethe
ansatz solution and discuss the equation of state and sound velocity
of a uniform Fermi gas at zero temperature. Then, by using LDA in
Sec. IV we determine the density distribution in the trapped environment.
In Sec. V, we describe the dynamics of trapped Fermi gases in terms
of 1D hydrodynamic equation. The behavior of low-lying collective
modes is obtained and discussed. Finally, we conclude in Sec. VI.

\section{Model Hamiltonian}

We consider a 1D multi-component Fermi gas with pseudo-spin $S=\left(\kappa-1\right)/2$,
where $\kappa$ ($\geq2$) is the number of components. The fermions
in different spin states repulsively interact with each other via
the \emph{same} short-range potential $g_{1D}\delta(x)$. The first-quantized
Hamiltonian with a total number of atoms $N=\sum_{l=1}^{\kappa}N_{l}$
(where $N_{l}$ is the number of fermions in the pseudo-spin state
$l$) for the system is

\begin{equation}
{\cal H}=\sum_{i=1}^{N}\left(-\frac{\hbar^{2}}{2m}\partial_{x_{i}}^{2}+V_{i}\right)+g_{1D}\sum_{i<j}\delta\left(x_{i}-x_{j}\right),\label{eq:Hami}
\end{equation}
where $V_{i}=m\omega^{2}x_{i}^{2}/2$ is the harmonic trapping potential
for the atom $i$ and $\omega$ is the trapping frequency. In general,
such a 1D Fermi system is created by loading a 3D cloud into a tight
2D optical lattice and separating it into a number of highly elongated
tubes. In each tube, it is convenient to express 1D coupling constant
$g_{1D}$ in terms of an effective 1D scattering length, 
\begin{equation}
g_{1D}=-\frac{2\hbar^{2}}{ma_{1D}},
\end{equation}
where the effective 1D scattering length $a_{1D}$ is related to the
3D scattering length $a_{3D}$ by the relation \cite{Olshanni1998,Bergeman2003,Astrakharchik2004}
, 
\begin{equation}
a_{1D}=-\frac{a_{\rho}^{2}}{a_{3D}}\left(1-\mathcal{A}\frac{a_{3D}}{a_{\rho}}\right)<0.
\end{equation}
Here, $a_{\rho}=\sqrt{\hbar/(m\omega_{\rho})}$ is the characteristic
oscillator length with transverse frequency $\omega_{\rho}$ determined
by optical lattice depth and 
\begin{equation}
\mathcal{A}\equiv-\frac{\zeta(1/2)}{\sqrt{2}}\simeq1.0326
\end{equation}
 is a constant.
\begin{quote}
\begin{figure}
\begin{centering}
\includegraphics[clip,width=0.8\textwidth]{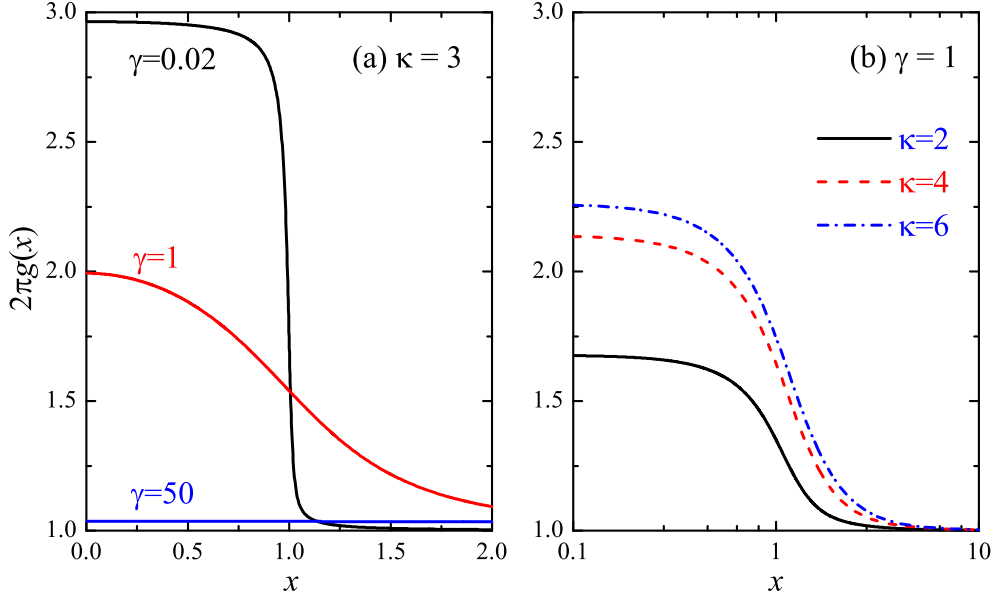} 
\par\end{centering}

\caption{(Color online) Dependence of the quasi-momentum distribution $g\left(x\right)$
on the interaction strength (a) and on the number of components (b).
At $x=0$, $g(x)$ increases with decreasing the interaction strength
or increasing the number of spin-components. While at large $x$,
$g(x)$ saturates to $1/2\pi$.}

\label{fig2} 
\end{figure}

\end{quote}

\section{Homogeneous equation of state}

Let us first address a 1D uniform multi-component Fermi gas with symmetric
inter-component interactions, i.e., there is no trapping potential
term $V_{i}$ in the Hamiltonian (\ref{eq:Hami}). In this case, the
model Hamiltonian is exactly soluble via the Bethe ansatz technique
\cite{Sutherland1968}. Focusing on the experiment at LENS, we assume
that each component has the same number of particles, i.e., $N_{l}=N/\kappa$
for $l=1,2,\cdots,\kappa=2S+1$. In free space, we measure the interactions
by a dimensionless coupling constant 
\begin{equation}
\gamma\equiv-\frac{mg_{1D}}{\hbar^{2}n}=\frac{2}{na_{1D}},
\end{equation}
where $n$ is the linear total number density \cite{amulti2}. The
ground state energy $E_{\hom}$ in the thermodynamic limit is given
by \cite{Sutherland1968,rmulti}, 
\begin{equation}
\frac{E_{\hom}}{L}=\frac{\hbar^{2}}{m}\left(\frac{n\gamma}{\lambda}\right)^{3}\int\limits _{0}^{1}x^{2}g\left(x\right)dx,\label{eq:energy}
\end{equation}
where

\begin{equation}
\lambda=2\gamma\int_{0}^{1}g\left(x\right)dx\label{eq: normalization}
\end{equation}
and the quasi-momentum distribution $g\left(x\right)$ with $x\geq0$
is determined by an integral equation

\begin{eqnarray}
g\left(x\right) & = & \frac{\kappa}{2\pi}-\frac{1}{\pi}\sum\limits _{l=1}^{\kappa-1}\int\limits _{1}^{\infty}g\left(x^{\prime}\right)dx^{\prime}\left[\frac{l\lambda}{\left(l\lambda\right)^{2}+\left(x-x^{\prime}\right)^{2}}\right.\nonumber \\
 &  & \left.+\frac{l\lambda}{\left(l\lambda\right)^{2}+\left(x+x^{\prime}\right)^{2}}\right].\label{eq: BetheAnsatz}
\end{eqnarray}
The above Bethe ansatz equations are very similar to those for attractive
interactions \cite{amulti1,amulti2,takahashi1,takahashi2}, but without
the contribution from $\kappa$-body bound cluster states. In Fig.
\ref{fig3}, we show the quasi-momentum distribution $g\left(x\right)$
for typical interaction parameters and the number of spin components,
obtained by solving the integral equation Eq. (\ref{eq: BetheAnsatz}),
together with Eq. (\ref{eq: normalization}).
\begin{quote}
\begin{figure}
\begin{centering}
\includegraphics[clip,width=0.6\textwidth]{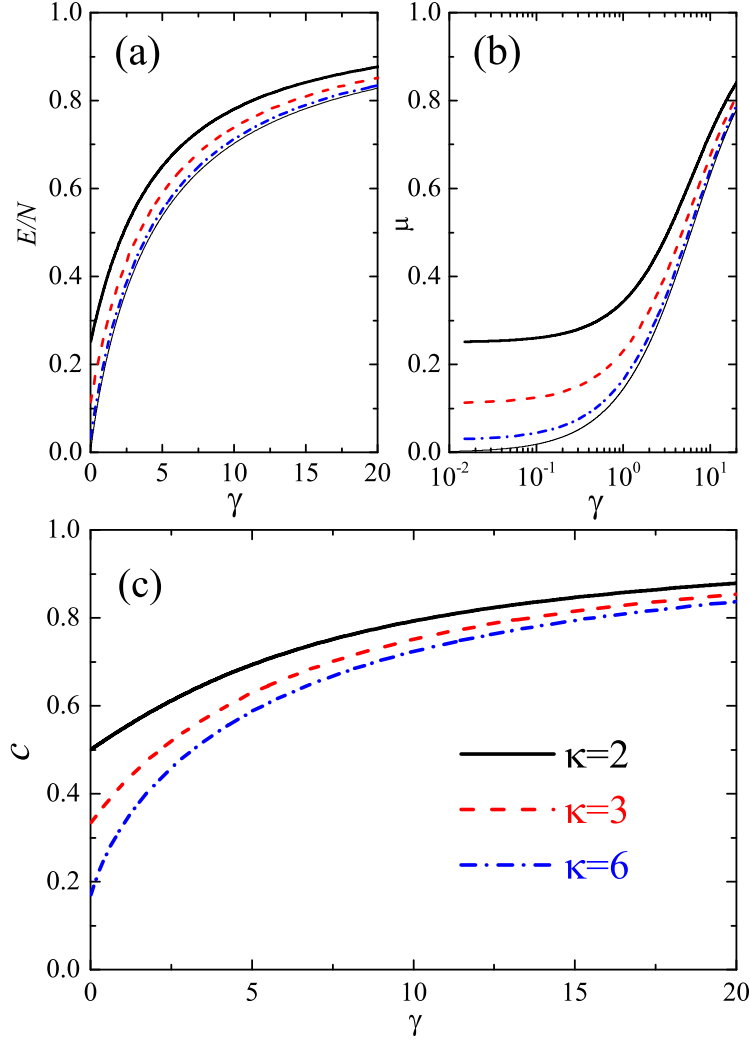} 
\par\end{centering}

\caption{(Color online) Dependence of the uniform ground state energy per particle
(a), the chemical potential (b), and the velocity of sound (c) on
the dimensionless coupling constant $\gamma$, at several number of
species as indicated. The energy per particle and sound velocity are
in units of the energy $\pi^{2}[\hbar^{2}n^{2}/(2m)]$ and the velocity
$\pi(\hbar n/m)$, respectively. Thin solid lines in (a) and (b) are
the results of a 1D spinless interacting Bose gas with the same density
$n$ and coupling constant $\gamma$. Note that in (b) for chemical
potential, $\gamma$ is shown in the logarithmic scale. Thus, the
saturation to the strongly interacting Tonk-Girardeau limit is not
obvious.}

\label{fig3} 
\end{figure}

\end{quote}
Once we obtain the ground state energy by using the Bethe ansatz technique,
we calculate the chemical potential by $\mu_{\hom}=\partial E_{\hom}/\partial N$
and the corresponding sound velocity by $c_{\hom}=\sqrt{n(\partial\mu_{\hom}/\partial n)/m}$.
For numerical purposes, it is convenient to rewrite these thermodynamic
quantities in a dimensionless form that depends on the coupling constant
$\gamma$ only, 
\begin{eqnarray}
\frac{E_{\hom}}{L} & \equiv & \frac{\hbar^{2}n^{3}}{2m}e\left(\gamma\right),\\
\mu_{\hom} & \equiv & \frac{\hbar^{2}n^{2}}{2m}\mu\left(\gamma\right),\\
c_{\hom} & \equiv & \frac{\hbar n}{m}c\left(\gamma\right).
\end{eqnarray}
These dimensionless functions are related by, 
\begin{eqnarray}
\mu\left(\gamma\right) & = & 3e\left(\gamma\right)-\gamma\frac{\partial e\left(\gamma\right)}{\partial\gamma},\\
c\left(\gamma\right) & = & \sqrt{\mu\left(\gamma\right)-\frac{\gamma}{2}\frac{\partial\mu\left(\gamma\right)}{\partial\gamma}}.
\end{eqnarray}
It is easy to see that for an ideal, non-interacting multi-component
Fermi gas, 
\begin{align}
\frac{E_{\hom}^{ideal}}{L} & =\frac{\hbar^{2}n^{3}}{2m}\left(\pi^{2}/3\kappa^{2}\right),\\
\mu_{\hom}^{ideal} & =\frac{\hbar^{2}n^{2}}{2m}\left(\pi^{2}/\kappa^{2}\right),
\end{align}
and 
\begin{equation}
c_{\hom}^{ideal}=\frac{\hbar n}{m}\left(\pi/\kappa\right).
\end{equation}

By numerically solving the integral Eqs. (\ref{eq:energy})-(\ref{eq: BetheAnsatz}),
we obtain the ground state energy per particle, chemical potential,
and sound velocity as a function of the dimensionless coupling constant
$\gamma$, as shown in Fig. \ref{fig3}. With increasing the coupling
constant, the energy, chemical potential and sound velocity increase
rapidly from the ideal gas results and finally saturate to the strongly
interacting Tonk-Girardeau gas limit, as one may anticipate. With
increasing the number of components $\kappa$, these thermodynamic
quantities decrease instead. It is interesting that for a sufficiently
large number of components, they approach to the results of a 1D repulsively
interacting spinless Bose gas with the same total density $n$ and
the coupling constant $\gamma=-mg_{1D}/(\hbar^{2}n)$, which are obtained
by solving the following integral equation \cite{takahashi1},
\begin{equation}
f\left(k\right)=\frac{1}{2\pi}+\frac{1}{\pi}\intop_{-1}^{+1}\frac{\lambda_{B}^{2}}{\lambda_{B}^{2}+\left(k-k'\right)}f\left(k'\right)dk',
\end{equation}
together with the normalization condition
\begin{equation}
\lambda_{B}=2\gamma\intop_{0}^{1}f\left(k\right)dk
\end{equation}
and the expression for the energy
\begin{equation}
\frac{E_{\hom}^{B}}{L}=\frac{\hbar^{2}}{m}\left(\frac{n\gamma}{\lambda_{B}}\right)^{3}\int\limits _{0}^{1}k^{2}f\left(k\right)dk.\label{eq:energyBosons}
\end{equation}
The equivalence between high-spin Fermi gas and spinless Bose gas,
namely high-spin bosonization, has been analytically shown by Yang
and You \cite{Yang2011}. This is an interesting counterpart of the
1D effective fermionization for strongly interacting particles in
one dimension \cite{Cazalilla2011}.

\section{Density distribution in harmonic traps}

\begin{figure}
\begin{centering}
\includegraphics[clip,width=0.6\textwidth]{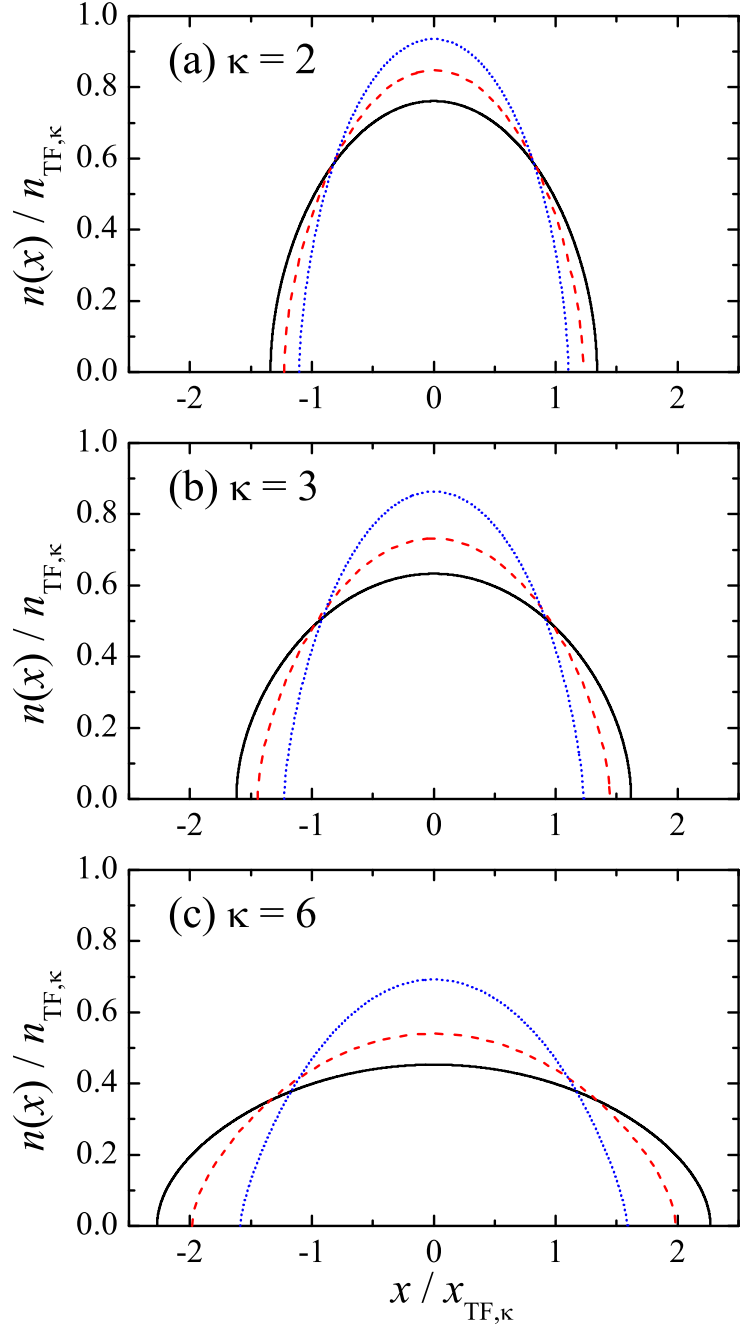} 
\par\end{centering}

\caption{(Color online) Density distributions of a 1D trapped multi-component
Fermi cloud at three interaction parameters $Na_{1D}^{2}/a_{ho}^{2}=0.1$
(black solid line), $1$ (red dashed line), and $10$ (blue dotted
line). The linear density and the coordinate are in units of the peak
density $n_{TF,\kappa}$ and Thomas-Fermi radius $x_{TF,\kappa}$
of an ideal Fermi gas, respectively.}

\label{fig4} 
\end{figure}

To make quantitative contact with the experiment at LENS \cite{YbLENS},
it is crucial to take into account the external harmonic trapping
potential $V(x)=m\omega^{2}x^{2}/2$, which is necessary to prevent
the fermions from escaping. We define a dimensionless parameter $\delta=Na_{1D}^{2}/a_{ho}^{2}$
to describe the interactions \cite{Astrakharchik2004,amulti2}. Here
$a_{ho}=\sqrt{\hbar/(m\omega)}$ is the characteristic oscillator
length in the axial direction. It is somehow counterintuitive that
$\delta\gg1$ corresponds to the weakly coupling limit, while $\delta\ll1$
corresponds to the strongly interacting regime. For a large number
of fermions, which is about $N\sim50$ experimentally \cite{YbLENS},
an efficient way to take the trap into account is by using the local
density approximation (LDA). Together with the exact homogeneous equation
of state of a 1D multi-component Fermi gas, this gives an asymptotically
exact results as long as $N\gg1$. The LDA amounts to determining
the chemical potential $\mu$ from the local equilibrium condition
\cite{Astrakharchik2004,ldh1d,hld1d},

\begin{equation}
\mu_{\hom}\left[n(x)\right]+\frac{1}{2}m\omega^{2}x^{2}=\mu_{g},
\end{equation}
under the normalization restriction $N=\int_{-x_{F}}^{+x_{F}}n\left(x\right)dx$,
where $n\left(x\right)$ is the total linear number density and is
nonzero inside a radius $x_{F}$. We have used the subscript ``$g$''
to distinguish the global chemical potential $\mu_{g}$ from the local
chemical potential $\mu_{\hom}$. Rewriting $\mu_{\hom}$ into the
dimensionless form $\mu[\gamma\left(x\right)]$ and $\gamma\left(x\right)=2/[n\left(x\right)a_{1D}]$,
we find that 
\begin{equation}
\frac{\hbar^{2}n^{2}\left(x\right)}{2m}\mu\left[\gamma\left(x\right)\right]+\frac{1}{2}m\omega^{2}x^{2}=\mu_{g}.
\end{equation}
We solve the above LDA equations numerically \cite{amulti2}.

\begin{figure}
\begin{centering}
\includegraphics[clip,width=0.8\textwidth]{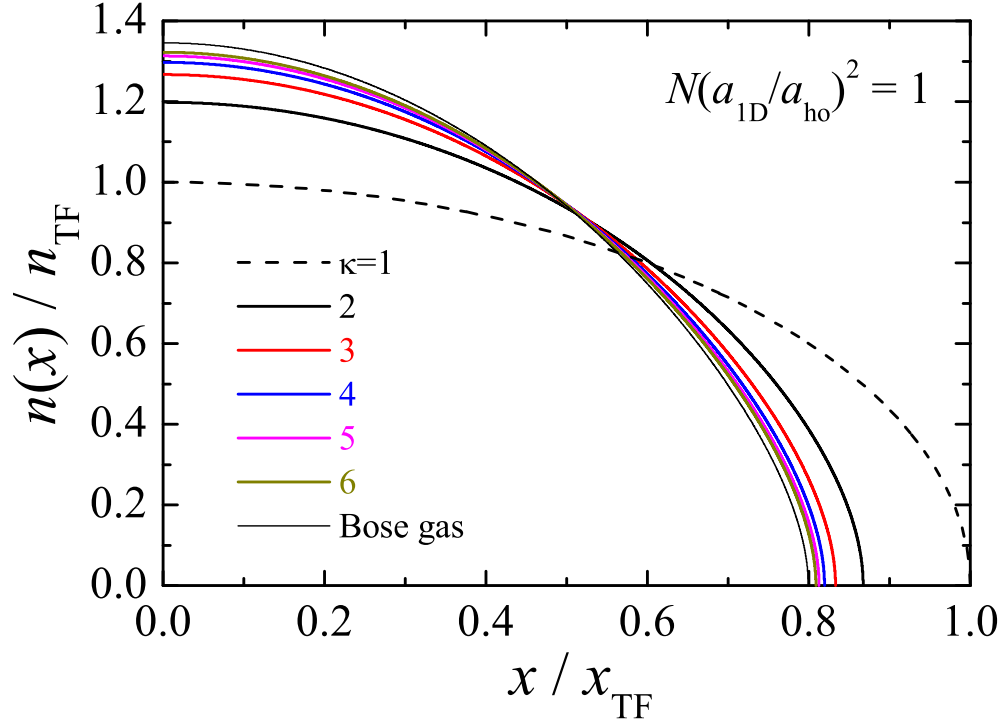} 
\par\end{centering}

\caption{(Color online) Evolution of density distributions with increasing
the number of components at the interaction parameter $Na_{1D}^{2}/a_{ho}^{2}=1$.}

\label{fig5} 
\end{figure}

Fig. \ref{fig4} reports the numerical results of the density distributions
at different number of component $\kappa$ and at three typical interaction
parameters $Na_{1D}^{2}/a_{ho}^{2}$. The linear density and the coordinate
are in units of the peak density $n_{TF,\kappa}=\sqrt{2N\kappa}/\left(\pi a_{ho}\right)$
and the Thomas-Fermi radius $x_{TF,\kappa}=\sqrt{2N/\kappa}a_{ho}$
of an ideal gas, respectively. With increasing interaction parameter
as shown in each panel, the density distribution changes from an ideal
gas distribution to a strongly interacting Tonks-Girardeau profile.
At the same interaction parameter, the density distribution become
flatter and broader as the number of components $\kappa$ increases. 

To show explicitly the effect of high-spin bosonization, in Fig. \ref{fig5}
we plot again the density distributions of $\kappa$-component Fermi
gas with varying $\kappa$ at a given dimensionless interaction parameter
$Na_{1D}^{2}/a_{ho}^{2}=1$, and compare them with the result of an
interacting spinless Bose gas, which is obtained by replacing $\mu_{\textrm{hom}}$
with $\mu_{\textrm{hom}}^{B}=\partial E_{\textrm{hom}}^{B}/\partial N$
(see Eq. (\ref{eq:energyBosons}) for $E_{\textrm{hom}}^{B}$) \cite{note}.
The profiles are shown in units of the peak density $n_{TF}=\sqrt{2N}/\left(\pi a_{ho}\right)$
and the Thomas-Fermi radius $x_{TF}=\sqrt{2N}a_{ho}$, for the purpose
of comparison. With increasing $\kappa$, the profiles converge quickly
to the density distribution of an interacting spinless Bose gas (thin
line), as a result of high-spin bosonization.

\section{Low-lying collective modes }

Experimentally, a useful way to characterize an interacting system
is to measure its low-lying collective excitations of density oscillations
\cite{Hu2004}. Quantitative calculations of the low-lying collective
excitations in traps can be based on the superfluid hydrodynamic description
of the dynamics of the 1D Fermi gas \cite{stringari}. In such a description,
the density $n\left(x,t\right)$ and the velocity field $v\left(x,t\right)$
satisfy the equation of continuity 
\begin{equation}
\frac{\partial n\left(x,t\right)}{\partial t}+\frac{\partial}{\partial x}\left[n\left(x,t\right)v\left(x,t\right)\right]=0,
\end{equation}
and the Euler equation 
\begin{equation}
m\frac{\partial v}{\partial t}{\bf +}\frac{\partial}{\partial x}\left[\mu_{\hom}\left(n\right)+V_{trap}\left(x\right)+\frac{1}{2}mv^{2}\right]=0.
\end{equation}
We consider the fluctuations of the density and the velocity field
about the equilibrium ground state , $\delta n\left(x,t\right)=$
$n\left(x,t\right)-n(x)$ and $\delta v\left(x,t\right)=v\left(x,t\right)-v(x)=v\left(x,t\right)$,
where $n(x)$ and $v(x)\equiv0$ are the equilibrium density profile
and velocity field. Linearizing the hydrodynamic equations, one finds
that \cite{stringari}, 
\begin{equation}
\frac{\partial^{2}}{\partial t^{2}}\delta n\left(x,t\right)=\frac{1}{m}\frac{\partial}{\partial x}\left\{ n\frac{\partial}{\partial x}\left[\frac{\partial\mu_{\hom}(n)}{\partial n}\delta n\left(x,t\right)\right]\right\} .
\end{equation}
The boundary condition requires that the current $J(x,t)=n(x)\delta v\left(x,t\right)$
should vanish identically at the Thomas-Fermi radius $x=\pm x_{TF}$.
Considering the $j$-th eigenmode with $\delta n\left(x,t\right)=\delta n\left(x\right)\exp\left[i\omega_{j}t\right]$
and removing the time-dependence, we end up with an eigenvalue problem,
\textit{i.e.}, 
\begin{equation}
\frac{1}{m}\frac{d}{dx}\left\{ n\frac{d}{dx}\left[\frac{\partial\mu_{\hom}(n)}{\partial n}\delta n\left(x\right)\right]\right\} +\omega_{j}^{2}\delta n\left(x\right)=0.\label{eq:hdyroeq}
\end{equation}

We note that the above hydrodynamic description is applicable in a
collisional regime characterized by the condition $N(1-\mathcal{T})\gg1$,
where $\mathcal{T}$ is the transmission coefficient for a 1D collision
of two fermionic atoms along the $x$-direction \cite{Olshanni1998}.
The calculation of the transmission coefficient $\mathcal{T}$ has
been given by Olshanni in his seminal work \cite{Olshanni1998}. By
estimating a collision wavevector $k_{z}\sim k_{F}$ and by using
the experimental parameters $N\sim100$ and $a_{\rho}/a_{3D}\simeq4.5$
at LENS \cite{YbLENS} , we find that $\mathcal{T}\sim0.3$ and $N(1-\mathcal{T})\sim70\gg1$.
Hence, the collisional regime is well reached in the recent LENS experiment
\cite{YbLENS}. 

\begin{figure}
\begin{centering}
\includegraphics[clip,width=0.8\textwidth]{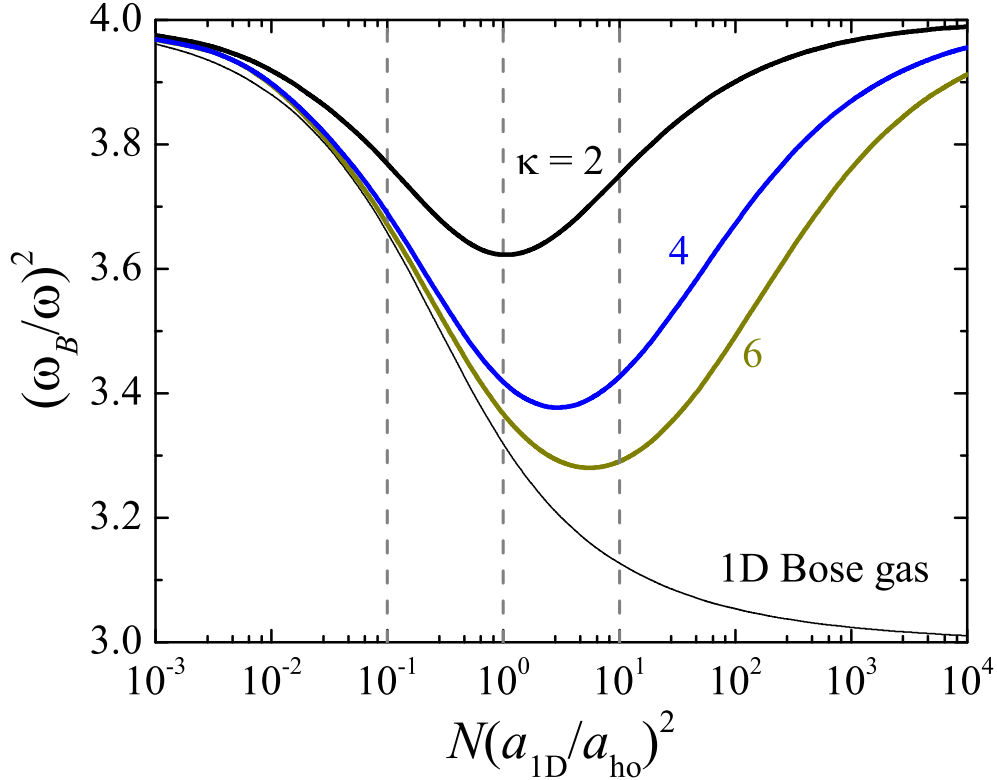} 
\par\end{centering}

\caption{(Color online) Breathing mode frequency as a function of the dimensionless
interaction parameter at different number of components. In the infinitely
large number of components, the mode frequency will approach to that
of a 1D interacting spinless Bose gas (thin line).}

\label{fig6} 
\end{figure}

To solve the hydrodynamic equation (\ref{eq:hdyroeq}), we use the
powerful multi-series-expansion method developed in Ref. \cite{amulti2}.
The resulting low-lying collective mode can be classified by the number
of nodes in its eigenfunction, \textit{i.e.}, the number index ``$j$''.
The lowest two modes with $j=1,2$ are the dipole and breathing (compressional)
modes, respectively, which can be excited separately by shifting the
trap center or modulating the harmonic trapping frequency. The dipole
mode is not affected by interactions according to Kohn's theorem,
and has an invariant frequency precisely at $\omega_{1}=\omega$.
Therefore, the mode frequency of the breathing mode provides the first
means to probe the non-trivial thermodynamics of our interacting system.

\begin{figure}
\begin{centering}
\includegraphics[clip,width=0.8\textwidth]{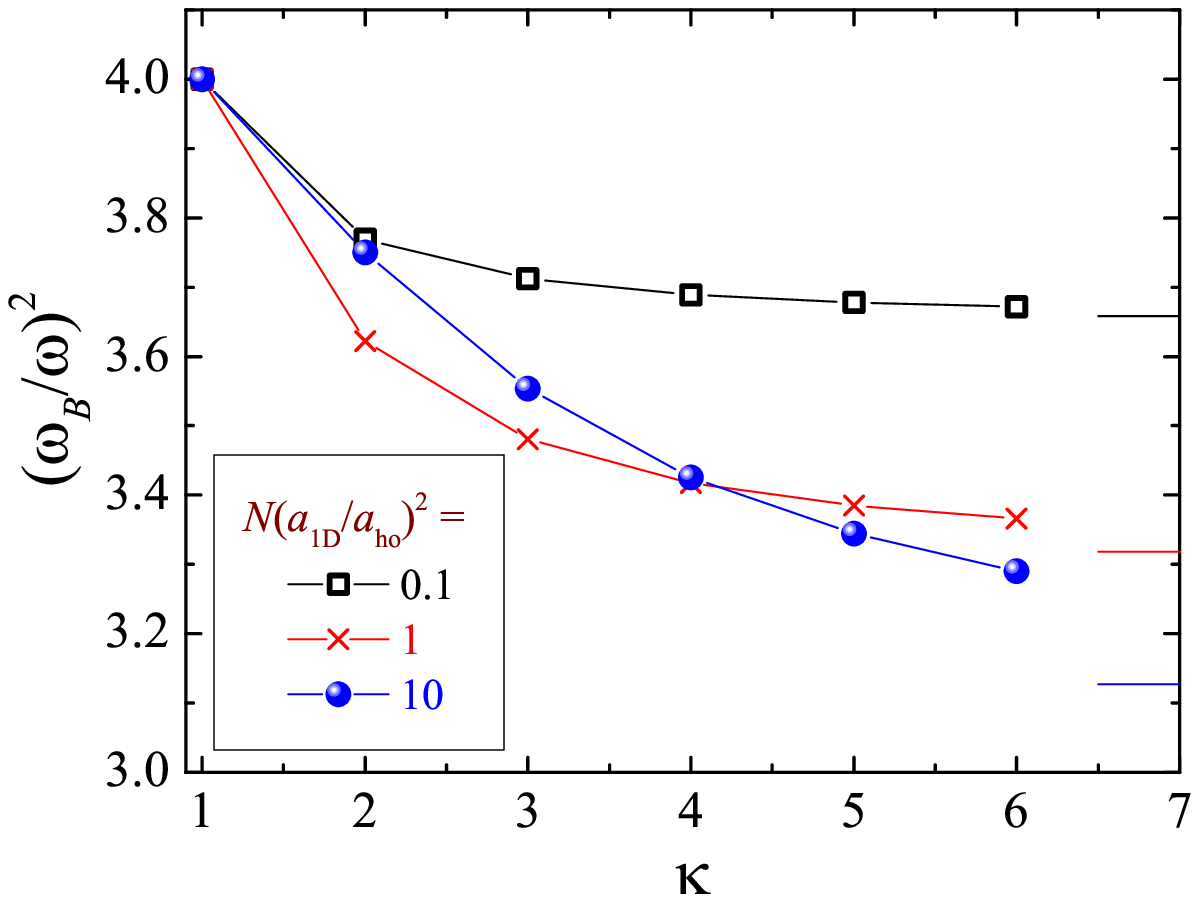} 
\par\end{centering}

\caption{(Color online) Illustration of high-spin bosonization in breathing
mode frequency. Here we show the evolution of breathing mode frequency
as a function of the number of components at the interaction strength
$Na_{1D}^{2}/a_{ho}^{2}=0.1$ (black empty squares), $1$ (red crosses),
and $10$ (blue solid circles). With increasing the number of components,
the breathing mode frequency approaches to the frequency of a 1D interacting
spinless Bose gas (thin lines).}

\label{fig7} 
\end{figure}

In Fig. \ref{fig6}, we show the breathing mode frequency as a function
of the dimensionless interaction parameter $\delta=Na_{1D}^{2}/a_{ho}^{2}$
at different number of components $\kappa$. In the weak coupling
limit ($\delta\gg1$) the cloud behaves likes an ideal Fermi gas,
whose breathing mode frequency is $2\omega$. In the strongly interacting
Tonks-Girardeau limit ($\delta\ll1$), the cloud is fermionized and
the mode frequency is again given by $2\omega$. Therefore, for a
given number of components, the breathing mode frequency exhibits
an interesting dip when the system crosses from the weak over to the
strong coupling regime. We note that for $\kappa=2$, such a dip structure
was predicted earlier by using a sum-rule approach \cite{Astrakharchik2004}.
With increasing the number of components $\kappa$, the mode frequency
decreases and finally approaches to that of a 1D interacting spinless
Bose gas. This high-spin bosonization behavior is highlighted in Fig.
\ref{fig7}. It should be noted that the Bose gas limit is very difficult
to reach in the weakly interacting regime when $\delta\gg1$. 

In Fig. \ref{fig1}, we compare our predictions for breathing mode
frequency with the experimental data reported by the LENS team \cite{YbLENS}.
The experiment was performed at an average interaction parameter $Na_{1D}^{2}/a_{ho}^{2}=0.44\pm0.08$.
The solid circles with error bars are the experimental results. The
empty squares with error bars are the theoretical results. The error
bar in the theoretical result is due to the uncertainty in the interaction
parameter $\Delta(Na_{1D}^{2}/a_{ho}^{2})=0.08$. The agreement between
theory and experiment, within a relative discrepancy of a few percents,
is impressive, as there is no any adjustable parameter. When the number
of component increases, both theoretical and experimental data approach
to the result of a 1D interacting spinless Bose gas, as indicated
by a thin horizontal line at the right part of the figure. This could
be viewed as an experimental proof of the high-spin bosonization phenomenon.

\begin{figure}
\begin{centering}
\includegraphics[clip,width=0.8\textwidth]{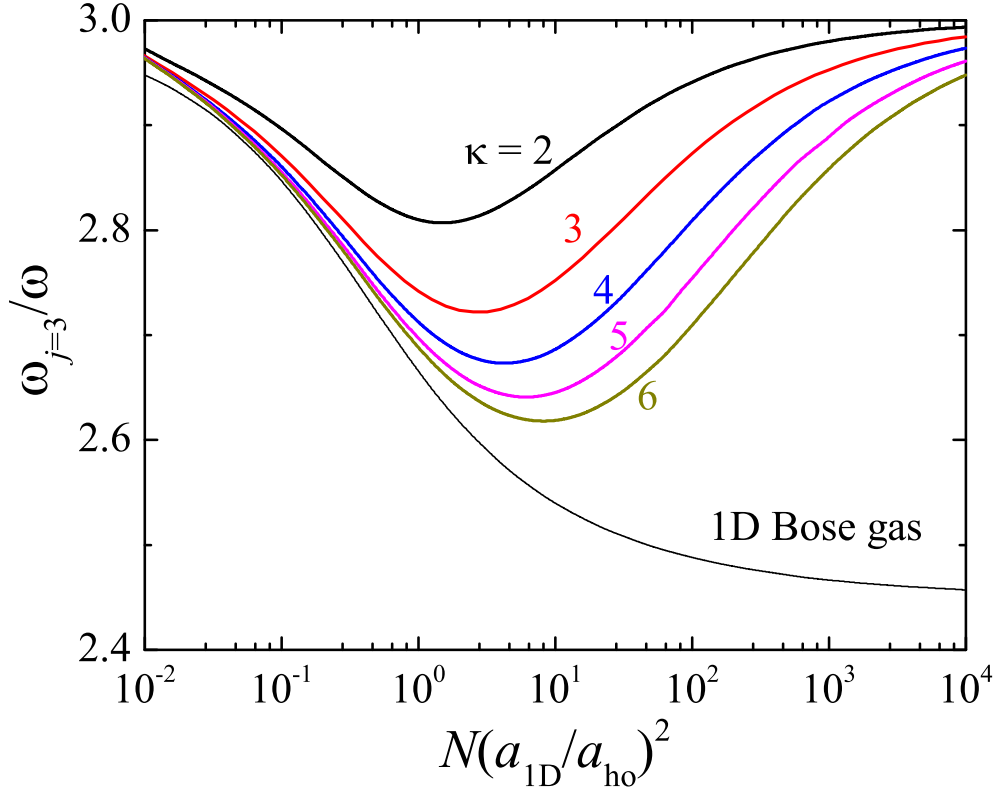} 
\par\end{centering}

\caption{(Color online) The 3rd mode frequency as a function of the dimensionless
interaction parameter at different number of components. In the infinitely
large number of components, the mode frequency will approach to that
of a 1D interacting spinless Bose gas (thin line).}

\label{fig8} 
\end{figure}

Qualitatively, collective modes with larger number of nodes ($j>2$)
exhibit the same feature as the breathing mode. An example is shown
in Fig. \ref{fig8}, for the frequency of the third collective mode.
Experimentally, however, these modes are more difficult to excite
and measure.

\section{Conclusion}

In summary, we have investigated the thermodynamics and collective
modes of 1D repulsively interacting Fermi gases with high-spin symmetry,
based on the exact Bethe ansatz technique beyond the mean-field method,
in a homogeneous environment. This has been extended to include a
harmonic trap, by using the local density approximation. The equation
of state of the system has been discussed in detail, as well as some
dynamical quantities, including the sound velocity and low-lying collective
modes. 

We have compared our collective mode prediction with a recent measurement
performed at LENS in a Fermi gas of $^{173}$Yb atoms, confined in
one dimension by using two-dimensional optical lattice. We have found
excellent quantitative agreement. In addition, we have predicted that
as the number of spin components $\kappa$ increases, the mode frequency
of the 1D repulsively interacting Femi gas approaches to that of a
1D interacting spinless Bose gas. This intriguing high-spin bosonization
phenomenon is qualitatively verified in the experiment in the regime
with an intermediate interaction strength.
\begin{acknowledgments}
This work was supported by the ARC Discovery Projects (Grant Nos.
FT130100815, DP140103231 and DP140100637) and NFRP-China (Grant No.
2011CB921502).\end{acknowledgments}


\begin{thebibliography}{10}
\bibitem{note}Similarly, all the results of a 1D interacting spinless
Bose gas, including the mode frequency that we will discuss later,
are obtained by replacing $\mu_{\textrm{hom}}$ with $\mu_{\textrm{hom}}^{B}$,
at the same total number of atoms $N$ and the coupling constant $Na_{1D}^{2}/a_{ho}^{2}$.
\end{thebibliography}
\end{document}